\documentclass[aps,pre,twocolumn,floatfix,nofootinbib,showpacs]{revtex4}
\usepackage{epsf,amsmath,amssymb,verbatim}
\begin{document}
\newcommand{\be}{\begin{equation}}
\newcommand{\ee}{\end{equation}}
\title{Structure and evolution of online social relationships:
Heterogeneity in warm discussions}
\author{K.-I. Goh$^*$, Y.-H. Eom$^{\dag}$, H.
Jeong$^{\dag}$, B. Kahng$^{*,\ddag}$, and D. Kim$^*$\\}
\affiliation{{$^*$School of Physics and Center for Theoretical
Physics, Seoul National University, Seoul 151-747, Korea}\\
{$^{\dag}$Department of Physics, Korea Advanced Institute of
Science and Technology, Daejon 305-701, Korea \\}
{$^{\ddag}$Center for Nonlinear Studies, Los Alamos National
Laboratory, Los Alamos, New Mexico 87545}}
\begin{abstract}
With the advancement in the information age, people are using
electronic media more frequently for communications, and social
relationships are also increasingly resorting to online channels.
While extensive studies on traditional social networks have been
carried out, little has been done on online social network. Here
we analyze the structure and evolution of online social
relationships by examining the temporal records of a bulletin
board system (BBS) in a university. The BBS dataset comprises of
1,908 boards, in which a total of 7,446 students participate. An
edge is assigned to each dialogue between two students, and it is
defined as the appearance of the name of a student in the from-
and to-field in each message. This yields a weighted network
between the communicating students with an unambiguous group
association of individuals. In contrast to a typical community
network, where intracommunities (intercommunities) are strongly
(weakly) tied, the BBS network contains hub members who
participate in many boards simultaneously but are strongly tied,
that is, they have a large degree and betweenness centrality and
provide communication channels between communities. On the other
hand, intracommunities are rather homogeneously and weakly
connected. Such a structure, which has never been empirically
characterized in the past, might provide a new perspective on
social opinion formation in this digital era.
\end{abstract}
 \maketitle \section{Introduction} With the advancement in
the information age, people are using electronic media for
communication more frequently, and social relationships between
people are also increasingly resorting to online communications.
For example, the advent of online bulletin board systems (BBS)
made it possible to develop a new type of online social
relationship and social consensus. Very similar to the Usenet
service, which was fairly popular during the earlier days of the
Internet, BBS is based on the communication between people sharing
common interests; the topic of interest is usually identified by
the board itself. People with common interests post messages on a
certain board and a response is conveyed by posting another
message, thereby forming a thread. Thus, a thread in the BBS
roughly represents a dialogue between people, and such a dialogue
constitutes the basic relationship among the people participating
in it. In the BBS, dialogues or discussions usually proceed with
little restriction on message writing and discrimination based on
personal information, thereby forming the so-called ``warm
discussions'' as described in psycho-sociology \cite{doise}.
Therefore, the pattern of such online social relationships may be
different from that of traditional social relationships based on
face-to-face contact or online communication involving exchange of
personal information, such as e-mail
transactions~\cite{bornholdt,huberman,dodds,eckmann,adamic} and
instant messaging \cite{messenger}. Thus, it would be interesting
to study the structure of online social relationship networks
constructed by people in warm discussions; this would be useful in
resolving diverse sociological and political issues and
understanding the manner in which social opinion is formed in the
digital era~\cite{axelrod,klemm,dodds_pnas,deffuant,weisbuch}.
Extensive studies on traditional social networks have been carried
out~\cite{milgram,granovetter,wasserman}; however, few studies
exist on online social networks. Here, we investigate the
structure of online social networks by studying BBS networks,
which are familiar to university students.

From the graph theoretical perspective, the BBS network offers
distinct features such as weighted and modular network structure.
Since the number of times a given pair of people exchange
dialogues can be counted explicitly, a weighted network is
naturally obtained~\cite{vespig_pnas}. Moreover, since people are
sharing a board corresponding to their common interests, BBS
provides an unambiguous way of defining modules or communities
\cite{gn}. This is unlike other examples of accessible protocols,
including the sibling/peer relationship in the online community
\cite{holme} and trackback in the blog system \cite{blog}. In
fact, the BBS network constructed by us differs in crucial aspects
from other affiliation networks such as the collaboration network
\cite{newman_coll} and student course registration network
\cite{course}. In these examples, the relationship between people
is not explicitly defined but is indicated indirectly by their
affiliation. Such an indirect definition generates several
cliques-completely connected subgroups-which may result in an
artifact particularly in the case of large-sized affiliations.
Thus, to obtain a network of people with explicit pairwise
interaction strength together with a distinct community definition
is crucial for an appropriate description of the social system.
The BBS network provides such ingredients.

\begin{figure}\centerline{\epsfxsize=.7 \linewidth \epsfbox{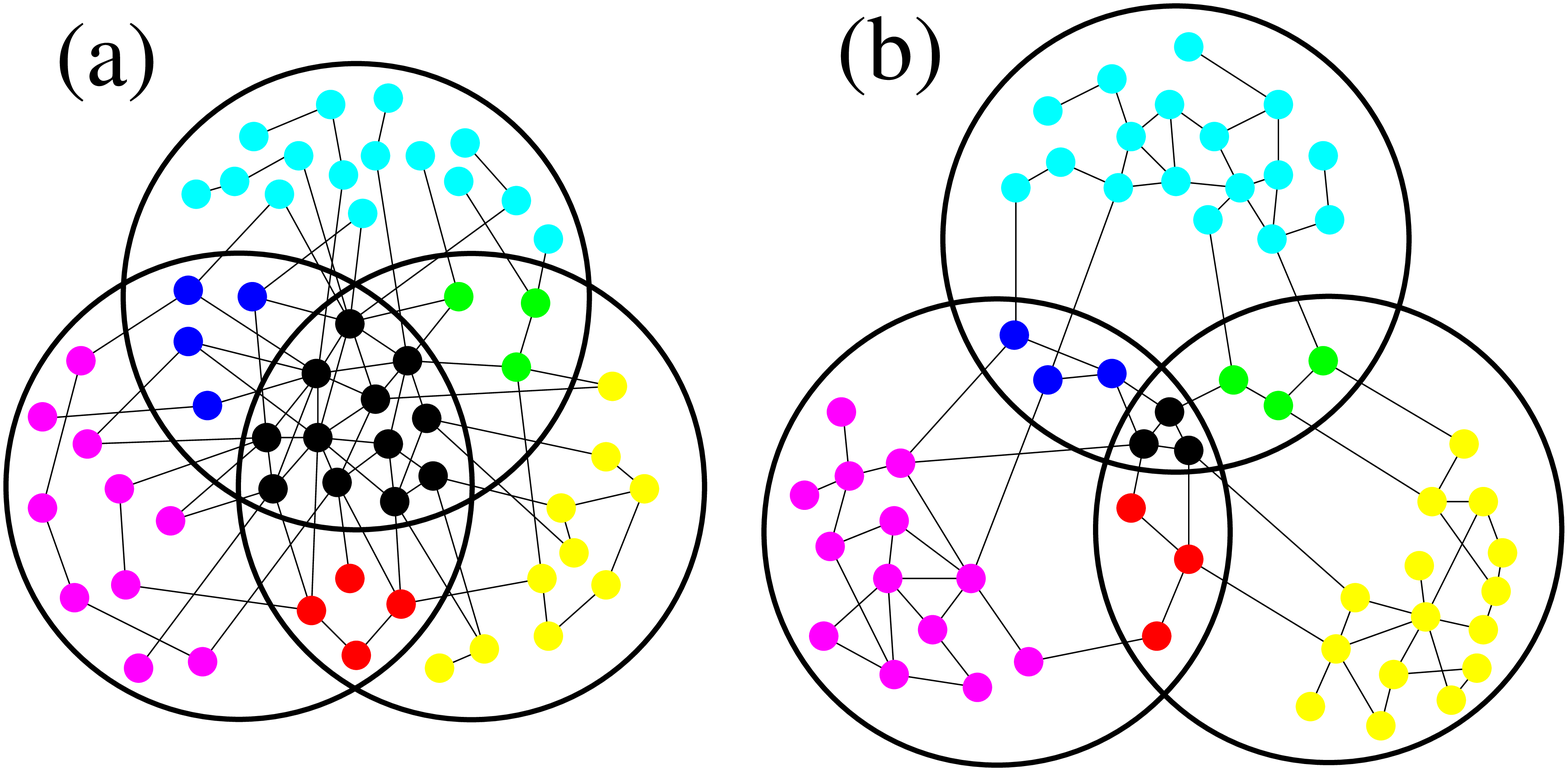}}
\caption{\sf Schematic network snapshots of the BBS network (a)
and traditional social network (b).} \end{figure}

\section{Conclusions and discussion}
The BBS network has interesting structural features and
implications. It contains hub members who participate in dialogues
across a large number of boards, thereby connecting one group of
people at one board to another group at a different board.
Further, their degrees, which are the numbers of people they have
exchanged dialogues with, are large, thereby influencing other
people throughout different communities. As a result, the hub
members act as weak ties in connecting different communities;
however, their links are strong during on actual activity. On the
other hand, intraboard connections are rather homogeneous in
degree. Such a network feature is in contrast to traditional
social networks maintained by the ties bridging disparate
communities, which tend to be weak~\cite{granovetter}. The
difference is schematically depicted in Fig.~1. In the BBS
network, the strength $s$, i.e., the total number of dialogues
each individual participates in has a nonlinear relationship with
the degree $k$ as $s\sim k^{1.4}$. This implies that the hub
members tend to post messages at considerably more frequently than
the other people with small degrees. The neutrality in the
assortative mixing is another feature of the BBS network compared
with the assortativity in traditional social networks. Such a
behavior may originate due to the absence of personal information
on the partner during online social communication. Thus, hub
members are democratic in their connections to the remaining
people, and they are indeed ``ubiquitous persons." Since the hub
members play a dominant role in providing communication channels
across different boards, it might be more efficient to use a
BBS-like online media for persuading people and drawing social
consensus than traditional social networks based on
person-to-person relationships. We attempt to understand the BBS
network from the perspective of a simple network model. In the
model, we take into account the empirical fact that the BBS
network contains groups of which size are inhomogeneous. In
addition, the link density of each group is not uniform, however
decreases with increasing group size, which has been usually
neglected in constructing model.

It would be interesting to implement the present work in the
context of a previous study involving a psycho-sociological
experiment on group discussions and the resulting
consensus~\cite{doise}, in which, group discussions are
distinguished into two types, ``warm'' and ``cold''. In the former
type, people express their thoughts freely without any
restriction, while in the latter, group discussions are restricted
by some constraint either explicitly or implicitly, for example,
the hierarchy in group members. The experimental study concludes
that the consensus measured after group discussions can be
different from that before the discussions depending on the type.
In the former, the consensus after discussions shifts to an
extreme opinions, while in the latter, it leads to a trade-off
average group consensus. From the perspective of the experiment,
we might state that the dialogues in the BBS are warm because no
restriction is imposed on posting messages and little information
on the personal background of the partner is provided. Thus, the
dialogues in the BBS may lead to radicalized consensus, violent
group behaviors, or imaginative and creative solutions to a given
issue. Since students still in the process of developing a value
system are vulnerable to negative influences, and have more
opportunities to be influenced by their peers through online
networks in this digital era than in the past, the proposed
network pattern we report here will be useful in guiding them in
the right direction. Moreover, the BBS network data will be
helpful in understanding the manner in which diverse opinions are
synchronized from the psycho-sociological perspective.

\section{BBS network}
We mainly examined the BBS system at the Korea Advanced Institute
of Science and Technology; it is named as {\tt loco.kaist.ac.kr}.
The characteristics of the network structure obtained from this
BBS system also appear in another system-{\tt bar.kaist.ac.kr}.
The data comprises records of all the threads posted  from March
9, 2000 to November 2, 2004, thus corresponding to a duration of
around three and a half years. As of November 2004, the system
comprised 1,908 boards with a total of 7,446 participating
students. In order to ensure privacy, we are only allowed to
access the information on ``from,'' ``to,'' the date of posting,
and the name of the board it was posted on, for each message.
Based on this information, we constructed the network between
students such that for each message, an edge was assigned between
two students appearing as ``from'' and ``to.'' Alternatively, an
arc (a directed edge) can be assigned for each message; however,
we found that the communications are largely reciprocal:
Approximately a half of the postings are accompanied by another
one with its from and to fields reversed, for example, a ``Re:''
message. Subsequently, we shall consider the network as undirected
for simplicity.

Our network construction naturally yields a weighted network in
which the weight $w_{ij}$ of the edge between two students $i$ and
$j$ is determined by the number of messages they exchanged during
the period. The detailed statistics of the BBS are listed in Table
I.

\begin{table}[p]
\caption{\sf Statistics of the BBS network as of November 2004.
The numbers in parentheses are the statistics for non-self
dialogues.}
\begin{tabular}{ll}
\hline \hline
Number of students $N$ & 7446 (7421) \\
Number of links $L$ & 103498 (103473) \\
Number of dialogues $W$ & 1299397 (1267292) \\
Number of boards $G$ & 1908 (1872) \\
Size of the largest cluster $N_1$ & 7350 \\
Average size of the boards $\bar{S}$ & 32.0 (32.6) \\
Average board memberships of a student $\bar{B}$ & 8.2 \\
Average path length $D$ & 3.3 \\
Mean degree $\langle k\rangle$ & 27.8 (27.9) \\
\hline \hline
\end{tabular}
\end{table}

\section{Structure of the BBS network}

\subsection{Student network}
The global snapshot of the student network in Fig.~1 reveals the
inhomogeneity among the students. The degree $k_i$ of a student
$i$, which is the number of students he/she has exchanged
dialogues with, is distributed according to a power law with an
exponent of around $-1$ followed by an exponential cutoff, as
shown in Fig.~2(a). This feature is similar to that of the
scientific collaboration network \cite{newman_coll}. The strength
$s_i$ of a student $i$ is the sum of the weight of each edge
attached to $i$. Therefore, $s_i=\sum_{j}^N a_{ij}w_{ij}$, where
$a_{ij}$ is the component of the adjacent matrix; its value is 1
if an edge is connected between vertices $i$ and $j$ and 0
otherwise. $w_{ij}$ is the weight of the edge between $i$ and $j$.
The strength and degree of a student exhibit a scaling behavior
$s(k)\sim k^{\beta}$ with $\beta\approx1.4$; however, the
fluctuation is quite strong, particularly for a small $k$
[Fig.~2(b)]. The strength distribution exhibits a behavior that is
similar to that of the degree distribution; however, the value of
the cutoff is larger[Fig.~2(a)]. The nonlinear relationship
between $s$ and $k$ implies that the hub members tend to post
messages at considerably more frequently than the other people, as
is evident in Table II.

\begin{table}[p]
\caption{\sf The fraction of the dialogues contributed by hub
members with a degree larger than 80 in the first ten longest
threads. The degree value of 80 is chosen approximately in
Fig.~2(a); beyond this degree, the power law for the degree
distribution fails.}
\begin{tabular}{lccccc}
\hline \hline
Rank~~~& Thread length~~~ & Number of dialogues ~~~ & Fraction \\
       &                  & contributed by hub members  & (\%) \\
\hline
1 & 229   & 181    & 79 \\
2 & 121   & 70     & 58  \\
3 & 92    & 92     &  100 \\
4 & 74    & 45     &  61 \\
5 & 67    & 16     &  24 \\
6 & 66    & 45     &  68 \\
7 & 65    & 27     &  41 \\
8 & 64    & 34     &  53 \\
9 & 54    & 54     & 100 \\
10 & 50   & 50    & 100 \\
\hline \hline
\end{tabular}
\end{table}

\begin{figure}\centerline{\epsfxsize=.7\linewidth \epsfbox{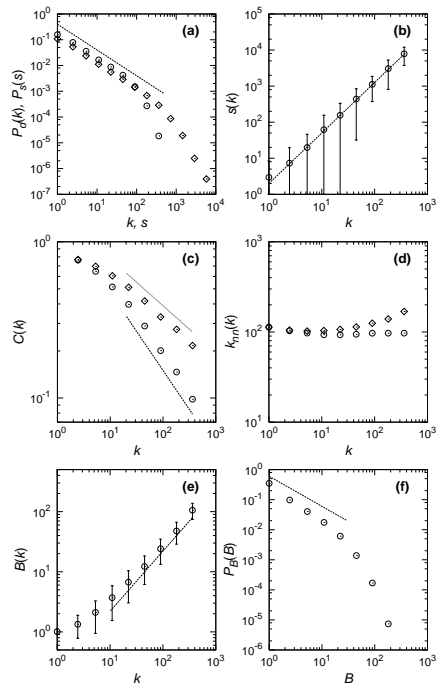}}
\caption{{\sf Structure of the BBS network.} (a) The degree
distribution $P_d(k)$ ($\circ$) and the strength distribution
$P_s(s)$ ($\diamond$) of the entire network. The straight line is
a guideline with a slope of $-1$. (b) The degree-strength scaling
relation $s(k)$. The straight line is a guideline with a slope of
$1.4$. (c) The clustering function $C(k)$ ($\circ$) and its
weighted version ($\diamond$). The straightlines are guidelines
with slope of $-0.5$ (lower) and $-0.3$ (upper), respectively. (d)
The average nearest-neighbor degree function $k_{\rm nn}(k)$ and
its weighted version ($\diamond$). (e) The correlation between the
degree and the membership number $B$. The dotted line is a
guideline with a slope of $1$. (f) The membership number
distribution of the vertices $P_B(B)$, where $B$ is the number of
boards that a student participates in. The straight line is a
guideline with a slope of $-1$.}
\end{figure}

Other standard measures of network topology are also obtained. The
local clustering coefficient $c_i$ is the local density of
transitive relationships, defined as the number of triangles
formed by its neighbors, cornered by itself, $i$, divided by the
maximum possible number of these, $k_i(k_i-1)/2$. The average of
$c_i$ over vertices with a given degree $k$ is referred to as the
clustering function $C(k)$. For the student network, $C(k)$ decays
as $\sim k^{-0.5}$ for large $k$, and its weighted version defined
in Ref.~\cite{vespig_pnas}\footnote{ In Ref.~\cite{vespig_pnas},
the local weighted clustering coefficient was defined as
$c_i^{(w)}=\sum_{j,h}(w_{ij}+w_{ih})a_{ij}a_{ih}a_{jh}/[2s_i(k_i-1)]$.
$C^{(w)}(k)$ is the average of $c_i^{(w)}$ over vertices with
degree $k$. The weighted average nearest-neighbors degree of
vertex $i$ was defined as $k_{{\rm nn},i}^{(w)}=\sum_{j=1}^N
a_{ij}w_{ij}k_j /s_i$. $k_{{\rm nn}}^{(w)}(k)$ is the average of
$k_{{\rm nn},i}^{(w)}$ over the vertices with degree $k$.} behaves
as $C^{(w)}(k)\sim k^{-0.3}$, as shown in Fig. 2(c). The
clustering coefficient $C$, which is the average of $c_i$ over all
vertices with $k>1$, is $\approx0.48$. This is one order of
magnitude greater than $C_{\textrm{random}}\approx 0.04$ of its
typical randomized counterpart with an identical degree sequence
\cite{maslov}. The average nearest-neighbor degree function
$k_{\rm nn}(k)$, which is defined by the average degree of the
neighbors of vertices of degree $k$, is almost flat for the
student network; nevertheless, its weighted version defined in
\cite{vespig_pnas} shows a slightly upward curvature for large $k$
(Fig. 2(d)). The assortativity coefficient \cite{assort} for the
binary network and the Spearman rank correlation of the degrees
are measured to be close to zero, as $r \approx 0.011$ and $r_{\rm
Spearman} \approx 0.024$, respectively. This almost neutral
mixing, which is in contrast to the common belief that social
networks are assortative, has also been observed in another online
social network \cite{holme}.

The number of boards that a student participates in is likely to
be larger for students with  a larger degree, as shown in
Fig.~2(e). Its distribution follows a skewed functional form in
Fig.~2(f). These results imply an important fact that a group of
people with a large degree tend to participate in diverse
dialogues on different boards and will play a dominant role in
drawing social consensus on diverse issues. Moreover, they work as
mediators between different groups in an online social community.

The betweenness centrality (BC) or load \cite{bc,bc_newman,load},
which is defined as the effective number of paths or packets
passing through a given vertex when every pair of vertices gives
and receives information, is also measured. The BC distribution
follows a power law with an exponent $\approx 2.2$, as shown in
Fig.~3(a) and the BC of a given vertex $\ell$ is strongly
correlated to its degree $k$ as $\ell \sim k^{1.6}$ as shown in
Fig.~3(b). This implies that the hub members have a large BC and
have a strong influence on the remaining people.

\begin{figure}\centerline{\epsfxsize=.7 \linewidth \epsfbox{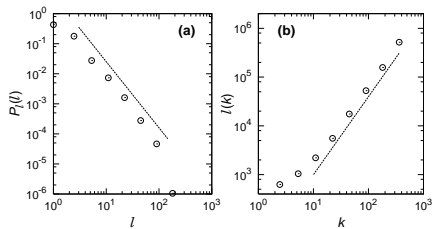}}
\caption{(a) The betweenness centrality (BC) distribution of the
BBS network. The dotted line is a guideline with a slope of
$-2.2$. (b) The relation between BC ($\ell$) and degree ($k$) of
the BBS network. The dotted line is a guideline with a slope of
$1.6$.} \end{figure}

In other words, the student network is extremely heterogeneous,
highly clustered, and yet, almost neutrally mixed, thereby
exhibiting a strong nonlinear relationship between the strength
and degree.

\subsection{Board network}
The procedure for constructing the board network is similar to the
usual projection method of the bipartite affiliation network. We
create a link between two boards if they share at least one common
member. In other words, each student participating in more than
one board contributes a complete subgraph---a clique--- to the
board network. Thus, the board network is the superposition of
cliques, each of which originates from the crossboard activities
of a student. Such crossboard activities will provide channels for
information transmission across the boards. In order to assign
meaningful weights to these channels, all the links in each clique
are assigned a weight that is equal to the inverse of the number
of vertices in that clique. In other words, the communication
channels created by the students posting on fewer boards are
stronger. Therefore, the weight of an edge between two boards
increases with the number of co-members; however, the
contributions of ``ubiquitous persons'' would only be moderate.
The strength of a board is the sum of the weights of its edges.
Such a strength distribution along with the degree distribution,
which does not account for the weight, is shown in Fig.~4(a). The
relation between the strength and degree is shown in Fig.~4(b).

\begin{figure}\centerline{\epsfxsize=.7 \linewidth \epsfbox{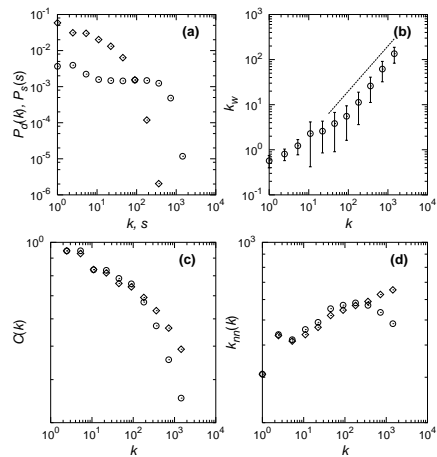}}
\caption{{\sf Structure of the board network.} (a) The degree
distribution $P_d(k)$ ($\circ$) and strength distribution $P_s(s)$
($\diamond$) of the board network. (b) The degree-strength
relation in the board network. The straight line is a guideline
with a slope of $1$. (c) The clustering function $C(k)$ ($\circ$)
and its weighted version ($\diamond$). (d) The average
nearest-neighbor degree function $k_{nn}(k)$ ($\circ$) and its
weighted version ($\diamond$).} \end{figure}

The board network is quite highly clustered with a clustering
coefficient of $\approx 0.61$, and the clustering function
decreases with $k$ [Fig.~4(c)]. However, it is noteworthy that
such a high clustering may result from the generation of cliques
by the projection procedure. Moreover, even the randomized board
network has a clustering coefficient as high as $\approx 0.48$.
The average nearest-neighbor degree initially increases with $k$
but decreases for larger $k$. However, its weighted version
increases monotonically with $k$, as shown in Fig.~4(d).

\section{Student network within a board}
Upon examining the networks within a board, we were presented with
a different scenario. As shown in Fig.~5(a), the degree
distributions of the student networks within the boards are rather
homogeneous. They exhibit a peak followed by an exponential tail,
which overall fits well into the Gamma distribution. Here, the
degree $k$ must be specified in further detail. Consider a case
where two students A and B on a given board who do not communicate
directly with each other. However, this communication between A
and B can occur on a different board. In this case, the two
students are regarded to be connected for the definition of degree
in Fig.~5(a). When such a pair is regarded to be disconnected, the
degree $k_0$ is redefined and its distribution exhibits fat tails,
as shown in Fig.~5(b); this was also observed in another BBS
system.

\begin{figure}\centerline{\epsfxsize=.7 \linewidth \epsfbox{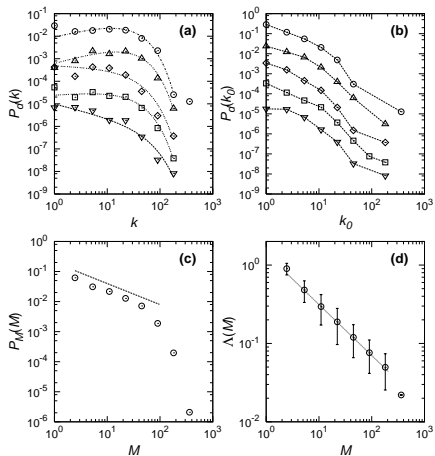}}
\caption{{\sf Properties of the board sub-network.} (a) The degree
distributions of subnetworks within the five largest boards.
Symbols used are ($\circ$), ($\triangle$), ($\diamond$), ($\Box$),
and ($\triangledown$) in the decreasing order of board size. The
fitted curves with the Gamma distribution
$k^{a-1}e^{-k/b}/[\Gamma(a)b^a]$ are shown. (b) The degree
distributions of subnetworks within the five largest boards with
degree redefined as discussed in the text. (c) The size
distribution of the boards $P_M(M)$. The straight line is a
guideline with a slope of $-0.7$. (d) The link density
$\Lambda(M)$ within a board as a function of its size $M$. The
straight line is a guideline with a slope of $-0.65$.}
\end{figure}

The size of the board, which denotes the number of students
posting messages on it, has a broad distribution [Fig.~5(c)]- a
power law followed by a rapidly decaying tail. The edge density
$\Lambda$ inside a given board scales with its size $M$ as
$\Lambda(M)\sim M^{-0.65}$, as shown in Fig.~5(d). Such a behavior
cannot be observed in the random sampling of populations of
different sizes, thereby indicating that the communications
between students are indeed strongly constrained within each board
rather than across them. Further, the power-law scaling behavior
suggests that the BBS network is organized in a self-similar
manner. From this result, it is evident that the usual projection
method involving the creation of cliques by bipartite affiliation
graphs cannot provide an appropriate description of the BBS
system. Moreover, such a size-dependent scaling of edge density
within groups has not been realized thus far in a simple model of
a clustered network~\cite{newman_cluster}.

\section{Evolution of the BBS network}

The daily record of the BBS network also allows us to examine the
temporal evolution of the network. The number of vertices
(students) $N$ grows exponentially after the transient period;
however, the continuously moderated growth rate appears to attain
a steady state [Fig.~6(a)]. Similar behavior is observed in the
case of the number of links $L$ and the number of dialogues $W$.
The number of boards $G$ grows at a rather steady rate over the
period.

\begin{figure}\centerline{\epsfxsize=.7 \linewidth \epsfbox{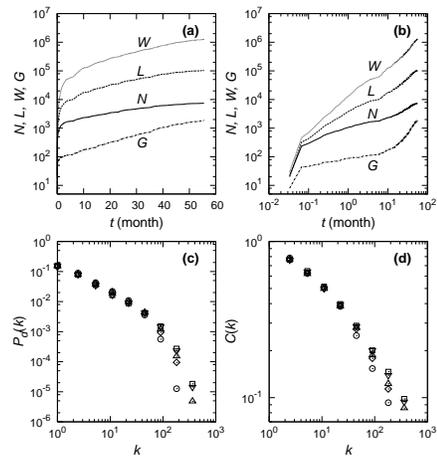}}
\caption{{\sf Evolution of the BBS network.} (a) The temporal
evolution of the number of students $N$ (solid), number of links
$L$ (dashed), total number of dialogues $W$ (dotted), and number
of boards $G$ (dot-dashed). (b) The same plot as (a) in the double
logarithmic scale. (c) The evolution of the degree distribution
$P_d(k)$ of the student network. The degree distribution for each
year is shown. The symbols ($\circ$), ($\diamond$), ($\triangle$),
and ($\triangledown$) correspond to each year from 2001 to 2004,
respectively, and ($\Box$) represents the final configuration. (d)
The clustering function $C(k)$ for each year. The same symbols as
those in (c) are used.} \end{figure}

Despite its continuous evolution, the structural properties of the
network seem to be in a stationary state. In other words, the
overall network characteristics such as the degree distribution
and clustering function achieve their forms in the initial period
(after $\sim$1 year), and do not change considerably with time, as
shown in Figs.~6(c) and (d). The crossover time scale of
approximately 1 year can also be observed in terms of the
evolution of the number of vertices $N$: Their growth patterns
change qualitatively after $\sim$10 months, as seen in Figs.~6(a)
and (b).

\section{Simple model}

Having identified the main statistical characteristics of the BBS
network, we attempt to understand them from the perspective of a
simple network model. First, we consider a simple extension of the
model of a clustered network introduced by
Newman~\cite{newman_cluster}. The original model of Newman is
specified with two fundamental probability distributions, $r_m$
and $s_M$. $r_m$ represents the probability that an individual
belongs to $m$ groups [$P_B(B)$ in our notation; (see Fig.~5(d))]
and $s_M$, the probability that the group size is $M$ [$P_M(M)$ in
our notation]. By assuming that the link density within the groups
is given by a constant parameter $p$, it is possible to obtain
several of formulae for the network structure using the generating
function method. For example, the degree distribution of the
network can be written as follows: \be P_d(k) =
\left.\frac{1}{k!}\frac{d^k}{dz^k}f_0[g_1(pz+q)]\right|_{z=0}, \ee
where $f_0(z)$ and $g_1(z)$ are appropriate generating functions
defined as $f_0(z)=\sum_{m=0}^{\infty}r_mz^m$ and $g_1(z)=\langle
M\rangle^{-1}\sum_{M=0}^{\infty}Ms_Mz^{M-1}$, and $q=1-p$.

However an obvious shortcoming of the model is that in real data,
the link densities are not uniform across the boards and they
strongly depend on the board size, as shown in Fig.~5(d). In fact,
by simply applying this model with the average link density
$p\approx0.3$ along with $r_m$ and $s_M$, directly measured from
the data, the degree distribution of the BBS network cannot be
reproduced. Therefore, we modify the model by allowing $p$ to vary
across the group, based on the empirical formula $\Lambda(M)\sim
M^{-0.65}$. Such a modification complicates the mathematical
formulae and they must be solved numerically. The resulting degree
distribution of the modified model along with that of the real
data is shown in Fig.~7. Although it is imperfect, the agreement
improved significantly. Thus, it is crucial to incorporate the
nonuniform link density into the realistic modeling of the BBS
network.

The manner in which the group size distribution, group membership
distribution, and group density scaling, which are the input
parameters of the model, achieve their present forms, as shown in
Figs.~5(c) and (d), is a topic for future study. \\

\begin{figure}\centerline{\epsfxsize=0.7\linewidth \epsfbox{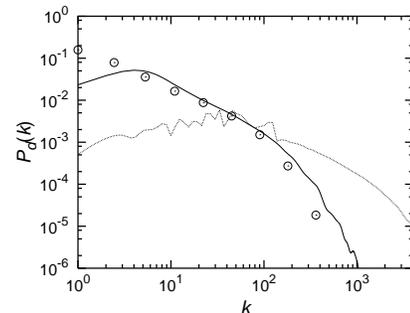}}
\caption{Comparison of the degree distributions of a simple model
of the BBS network of Newman (dotted) and its modification
(solid), with that of the real network (circle).} \end{figure}

This work was supported by KRF Grant No. R14-2002-059-010000-0 of
the ABRL program funded by the Korean government MOEHRD and
the CNS research fellowship from SNU (BK).\\

\end{document}